\documentclass[prl,superscriptaddress,preprint,endfloats,showpacs]{revtex4}

\newcommand {\CA}{Cd$_3$As$_2$}
\newcommand {\ZA}{Zn$_3$As$_2$}
\newcommand {\NB}{Na$_3$Bi}
\newcommand {\BS}{Bi$_{1-x}$Sb$_x$}
\newcommand {\BTSS}{BiTl(S$_{1-x}$Se$_x$)$_2$}
\newcommand {\CZA}{(Cd$_{1-x}$Zn$_x$)$_3$As$_2$}

\newcommand {\SN}{Si$_3$N$_4$}
\newcommand {\STO}{SrTiO$_3$}

\usepackage{graphicx}
\usepackage{bm} 
\usepackage{pifont}
\usepackage{amsmath, amsthm, amssymb}
\usepackage{dcolumn} 
\usepackage{color}
\def\vector#1{\mbox{\boldmath $#1$}}
\begin{document}
\title{
Negative magnetoresistance suppressed through topological phase transition in (Cd$_{1-x}$Zn$_x$)$_3$As$_2$ films
}
\author{S. Nishihaya}
\affiliation{Department of Applied Physics and Quantum-Phase Electronics Center (QPEC), University of Tokyo, Tokyo 113-8656, Japan}
\author{M. Uchida}
\email[Author to whom correspondence should be addressed: ]{uchida@ap.t.u-tokyo.ac.jp}
\affiliation{Department of Applied Physics and Quantum-Phase Electronics Center (QPEC), University of Tokyo, Tokyo 113-8656, Japan}
\author{Y. Nakazawa}
\affiliation{Department of Applied Physics and Quantum-Phase Electronics Center (QPEC), University of Tokyo, Tokyo 113-8656, Japan}
\author{K. Akiba}
\affiliation{Institute of Solid State Physics (ISSP), University of Tokyo, Chiba 277-0882, Japan}
\author{M. Kriener}
\affiliation{RIKEN Center for Emergent Matter Science (CEMS), Wako 351-0198, Japan}
\author{Y. Kozuka}
\affiliation{Department of Applied Physics and Quantum-Phase Electronics Center (QPEC), University of Tokyo, Tokyo 113-8656, Japan}
\author{A. Miyake}
\affiliation{Institute of Solid State Physics (ISSP), University of Tokyo, Chiba 277-0882, Japan}
\author{Y. Taguchi}
\affiliation{RIKEN Center for Emergent Matter Science (CEMS), Wako 351-0198, Japan}
\author{M. Tokunaga}
\affiliation{Institute of Solid State Physics (ISSP), University of Tokyo, Chiba 277-0882, Japan}
\author{M. Kawasaki}
\affiliation{Department of Applied Physics and Quantum-Phase Electronics Center (QPEC), University of Tokyo, Tokyo 113-8656, Japan}
\affiliation{RIKEN Center for Emergent Matter Science (CEMS), Wako 351-0198, Japan}

\begin{abstract}
The newly discovered topological Dirac semimetals host the possibilities of various topological phase transitions through the control of spin-orbit coupling as well as symmetries and dimensionalities. Here, we report a magnetotransport study of high-mobility {\CZA} films, where the topological Dirac semimetal phase can be turned into a trivial insulator via chemical substitution. By high-field measurements with a Hall-bar geometry, magnetoresistance components ascribed to the chiral charge pumping have been distinguished from other extrinsic effects. The negative magnetoresistance exhibits a clear suppression upon Zn doping, reflecting decreasing Berry curvature of the band structure as the topological phase transition is induced by reducing the spin-orbit coupling.
\end{abstract}
\pacs{75.47.-m, 72.15.Lh, 73.50.-h}
\maketitle

Topological Dirac semimetal (DSM), where pairs of three-dimensional band touching points (Dirac points) are stabilized by band inversion and rotational symmetry \cite{topo3}, has been the first gapless yet topological system experimentally found in solids. This topological phase, realized in such compounds as {\CA} \cite{CdAst, CdAsARPES1, CdAsARPES2} and {\NB} \cite{NaBit, NaBiARPES}, can be regarded as an intermediate state adjacent to other exotic topological phases, and thus it serves as a useful platform for pursuing topological phase transitions \cite{CdAst, NaBit, CdAskwsk}. Tuning the dimensionality to realize two-dimensional quantized conduction in {\CA} thin films \cite{CdAskwsk} is one example. Another possible way for driving phase transitions is to control the strength of spin-orbit coupling (SOC) to modulate the band topology. While the effectiveness of the SOC modulation in controlling the topological phases has been demonstrated in topological insulators such as {\BS} \cite{TPT1} and {\BTSS} \cite{TPT2}, the tunability of the band structure through SOC is of crucial importance especially in DSM because its topological nature directly originates from the three-dimensional Dirac points (i.e., magnetic monopoles). In addition to the topological phase transitions \cite{NaBit}, shifts of the monopoles in the momentum space can also lead to a modulation of topological transport phenomena in DSM \cite{CAexp1, CAexp2, CAexp3, CAexp4, Topotr1, Topotr2, Topotr3, Topotr4}, providing more degrees of freedom for investigating and controlling DSM. 

For the DSM compound {\CA}, this approach is possible by the chemical substitution of Zn at the Cd site. {\ZA}, which forms a complete solid solution with {\CA} \cite{CZA1}, is a trivial $p$-type semiconductor with an ordinary band ordering. Doping Zn can lift the band inversion in {\CA}, leading to a transition from DSM to a trivial insulator \cite{CZA2, CZA3}. One of the transport properties of particular interest in such a topological phase transition is negative magnetoresistance (MR) induced by the so-called chiral anomaly \cite{CA1, CA2, CA3}. In a three-dimensional gapless system, the lowest Landau levels (LLLs) form one-dimensional zero modes along the direction of the applied magnetic field \vector{B}. In the presence of electric field \vector{E}, the LLL pumps charges from one valley into another valley with opposite chirality $\chi\ (=\pm 1)$ at the rate of $\chi\frac {e^{2}}{4\pi^{2}\hbar^{2}} (\vector{E} \cdot \vector{B})$. Here, $e$ is the elementary charge and $\hbar$ is the Plank constant divided by 2$\pi$. The charge pumping between the valleys leads to the emergence of an additional charge current and hence negative MR \cite{CAexp1, CAexp2, CAexp3, CAexp4}. In addition to the above quantum limit picture, it has also been proved that the same negative magnetoresistance can be derived semiclassically at a weak field limit, where the non-zero Berry phase is responsible for the charge pumping under the parallel $\vector{E}$ and $\vector{B}$ \cite{CA3}.

Experimentally, however, negative MR can be caused not only by the chiral anomaly but also by other extrinsic origins such as the current jetting effect \cite{CJ}, leaving the interpretation of the observed negative MR difficult and elusive. From this viewpoint, it is highly meaningful to investigate the magnetotransport \color{black}through the topological phase transition process of the {\CZA} alloy system. Since the origin of the chiral anomaly in DSM lies on the existence of the three-dimensional Dirac points, this characteristic electromagnetic effect is expected to exhibit a clear response to the phase transition. As depicted in Fig. 1, the charge pumping under $\vector{E}$ parallel to $\vector{B}$ condition is practically governed by the back-scattering time $\tau_{\rm v}$ between the two valleys \cite{CA2, CA3}. Therefore, as the distance between the valleys is reduced with approaching the phase transition, the charge pumping is suppressed due to increased inter-valley scattering.

In this work, we study the magnetotransport of {\CZA} thin films at high fields, to capture the transport manifestation of the expected topological phase transition process. We demonstrate that the negative MR observed in the DSM phase of {\CA} exhibits a clear suppression upon Zn doping, suggesting that the DSM phase is successfully modulated through the topological phase transition which is induced by controlling the SOC strength.

\begin{figure}
\begin{center}
\includegraphics[width=12.5cm]{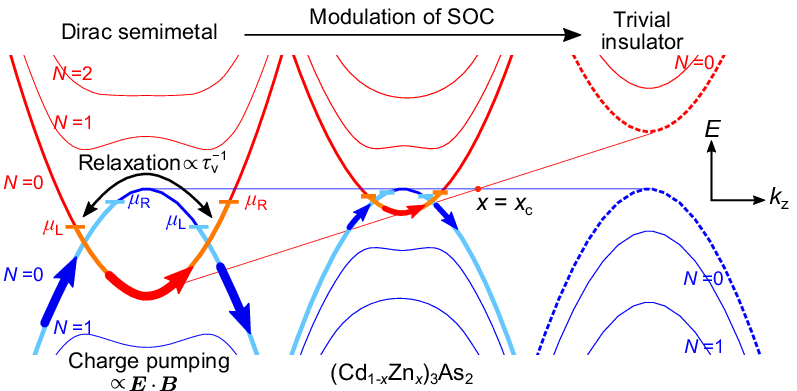}
\caption{
(color online).
Schematics of the topological phase transition from the Dirac semimetal (DSM) to a trivial insulator through chemical modulation of the spin-orbit coupling (SOC) in {\CZA}. The band inversion disappears at the critical composition $x = x_{\rm c}$. In the DSM phase, the lowest Landau levels (LLLs) form one-dimensional chiral modes to pump charges from one valley with chirality $\chi = +1(-1)$ to another valley with $\chi = -1(+1)$ under the application of parallel electric and magnetic fields $\vector{E}\cdot\vector{B} >0(<0)$. $\mu_{\rm R}(\mu_{\rm L})$ denotes the chemical potential of the charges with $\chi = +1(-1)$. In the steady state, there exists inter-valley scattering to back-scatter the pumped charges. As the band inverted region shrinks by Zn doping, the inter-valley scatterings occur more frequently, leading to the suppression of the chiral-anomaly-induced conduction.
}
\label{fig1}
\end{center}
\end{figure}

{\CZA} thin films with a Zn concentration $x$ ranging from 0 to 0.21 were fabricated on {\STO} (100) substrates by a combination of pulsed laser deposition and thermal treatment, following the same procedure described in Ref. \cite{CdAskwsk, CdAskwsk2}\color{black}. 
{\CA} and {\ZA} layers with appropriate thickness ratios are first deposited, which are then covered by MgO and {\SN} capping layers and annealed at 600 C$^{\circ}$. The annealing process promotes the crystallization as well as the formation of a solid solution of the two arsenide layers (Fig. 2(a)). Figure 2(b) summarizes the x-ray diffraction patterns of the {\CZA} films. The {\CZA} alloy films are confirmed to have a (112) oriented single phase. The clear peak shifts indicate that the lattice constant systematically shrinks upon Zn doping. In {\CZA}, the lattice parameters are known to follow Vegard's law \cite{CZA1,CZA3}. Thus, the Zn concentration $x$ can be estimated from the lattice constant. Figure 2(c) shows the (112) lattice spacing $d_{112}$ and the estimated $x$ values as a function of deposited {\CA}/{\ZA} thickness ratio $x_{\rm i}$. The estimated $x$ is in a good agreement with the designed value $x_{\rm i}$, indicating excellent controllability and spatial uniformity of the Zn doping. Because of the symmetry difference between the film and substrate, there exist inplane domains where the hexagonal patterns of the {\CZA} (112) plane are $90^{\circ}$ rotated with respect to the {\STO} (100) plane \cite{CdAskwsk, CdAskwsk2}. The typical domain size is over 10 $\mu$m which is comparable to the channel size of 60 $\mu$m \cite{CdAskwsk, CdAskwsk2}. The films are designed to be thick enough ($85 \sim 100$ nm) to maintain the three-dimensional Fermi surface.

Making a solid solution with $p$-type {\ZA} causes not only the modulation of SOC strength, but also the compensation of the defect-induced electron density in {\CA}. While all the samples exhibit a metallic behavior in the temperature dependence of the resistivity (Fig. 2(d)), the electron density is systematically suppressed by Zn doping as shown in Fig. 2(e). The mobility is kept over 10$^{4}$ cm$^{2}$/Vs (Fig. 2(f)), which is similar to those found in {\CZA} bulk single crystals reported in Ref. \cite{CZA3}.

All the transport properties were measured using the samples patterned into a Hall-bar shape (the inset of Fig. 2(e)). In this way, we have carefully avoided inhomogeneous current flow in order to exclude the current-jetting effect \cite{CJ} on the MR measurements.

\begin{figure*}
\begin{center}
\includegraphics[width=16.0cm]{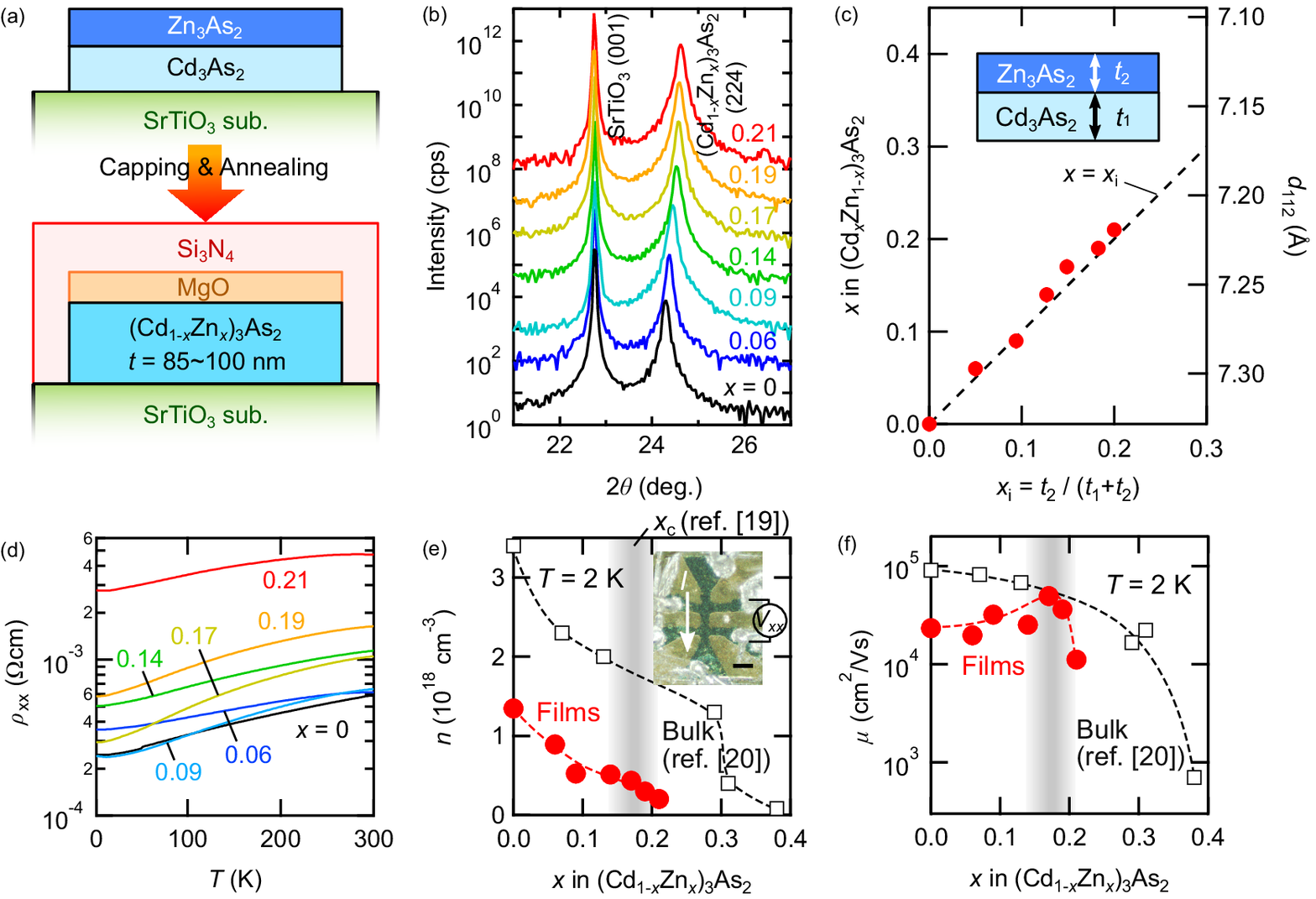}
\caption{
(color online). (a) Schematic illustration of the capping and annealing process. (b) X-ray diffraction patterns of {\CZA} films around the (224) peak (vertically shifted for clarity). (c) (112) lattice spacing $d_{112}$ and estimated Zn concentration $x$ as a function of {\CA}/{\ZA} thickness ratio $x_{\rm i}$. (d) Temperature dependence of resistivity for the (112) oriented {\CZA} films. All the samples exhibit a metallic bdehavior. (e) Carrier density and (f) mobility of the {\CZA} films as a function of Zn concentration $x$. $x_{\rm c}$ is estimated to be around 0.17 based on Ref. \cite{CZA2}. The mobility in these films is close to those reported for bulk single crystals (open square, Ref. \cite{CZA3}). The inset in (e) shows a photo of the film sample patterned into a Hall-bar shape for the magnetoresistance measurements. The scale bar is 100 $\mu$m.  
}
\label{fig2}
\end{center}
\end{figure*}

MR of the {\CZA} thin films was measured under magnetic fields up to 55 T. Figure 3(a) presents the data of the $x = 0$ sample when the field is rotated from the out-of-plane direction to the in-plane current direction. When the field is aligned to the current ($\theta = 0$), a clear negative MR is observed up to high fields. The appearance of quantum oscillations with the in-plane magnetic field ensures the three-dimensionality of the electronic state in the film sample.

While appearance of negative MR in topological Dirac semimetals including {\CA} has been reported in previous studies \cite{CAexp1, CAexp2, CAexp3, CAexp4, CAexp5}, it requires careful assessments to identify the underlying origins. Aside from the current jetting effect \cite{CJ} which we have carefully avoided by the Hall-bar measurements, there are many other origins that may cause appearance of the negative MR; namely classical size effect \cite{size1, size2}, conductivity fluctuation \cite{CAexp5}, and gap-closing of LLLs in band-inverted materials \cite{NMR_LLL}. Below we discuss these scenarios to show that they can be excluded from the possible origins of the negative NR observed in the {\CZA} films.

The first is the classical size effect which often appears in nano-strcuture samples \cite{size1, size2}. When the mean free path $l_{e}$ of charge carries is limited by the channel width $W$, the zero-field resistivity increases by a certain factor ($\mathit{\Delta} \rho \sim \frac{l_{e}}{2W} \rho_0$) from the original bulk value $\rho_0$ due to back-scattering at the channel boundary \cite{size1}. Under magnetic fields, the boundary scatterings are then suppressed in the presence of the cyclotron motion of charge carriers, leading to the decrease of the resistivity. 
In our film samples, the mean free path ($l_{e} = 50\sim 100$ nm) is comparable to the film thickness and the negative MR due to the size effect is indeed observed especially when the in-plane transverse MR is measured as shown in Fig. 3(b) (see also the Supplemental Material \cite{supplemental}). However, such an effect becomes saturated at the field where the cyclotron diameter becomes smaller than the film thickness  and thus it is limited to low fields (the shaded region in Fig. 3(b) and 3(c))\cite{supplemental, size1, size2}. The observed negative MR in the {\CA} film, on the other hand, persists even up to the field where quantum oscillations are well developed (Fig. 3(c)), indicating that the negative component at such a high field cannot be explained by the classical size effect.  

\color{black}
\begin{figure}
\begin{center}
\includegraphics[width=16.0cm]{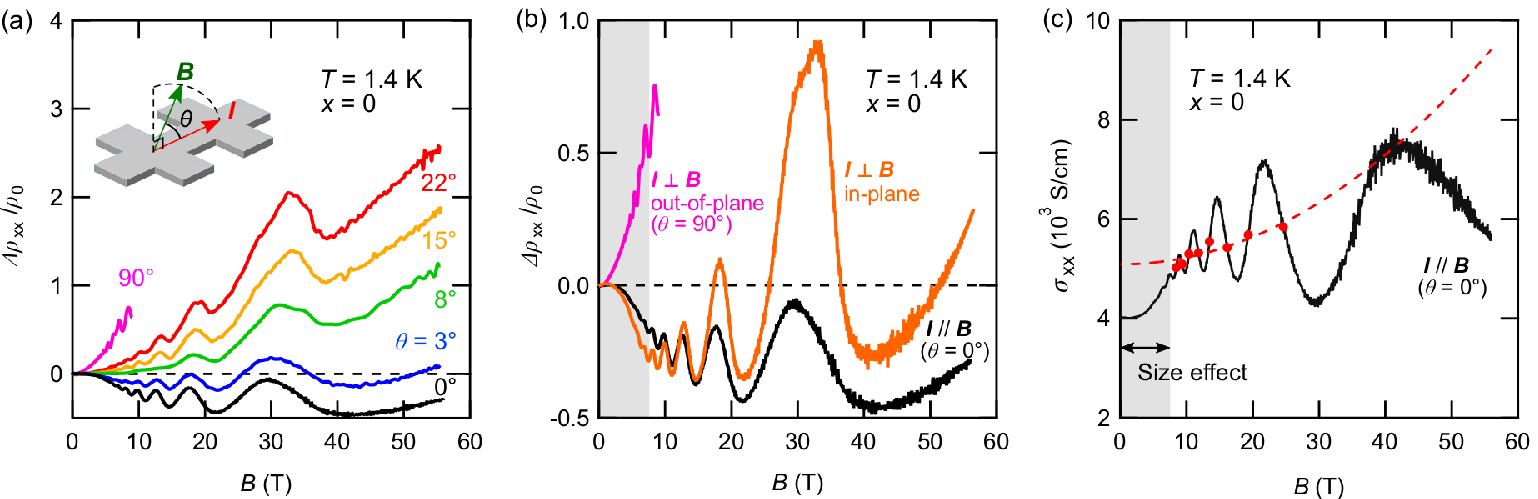}
\caption{
(color online).
(a) Magnetoresistance (MR) at $T$ = 1.4 K with the magnetic field tilted from the out-of-plane to the current direction. $\theta$ denotes the angle between field $B$ and current $I$. Negative longitudinal MR is observed up to high fields at $\theta = 0$. The existence of the quantum oscillations at $\theta = 0$ ensures the three-dimensionality of the Fermi surface. (b) In-plane transverse MR (denoted as $I\perp B$ in-plane) shown along with longitudinal MR ($I\parallel B$) and out-of-plane transverse MR ($I\perp B$ out-of-plane). The drop of in-plane transverse MR in the lower field (the shaded region) indicates the existence of classical size effect. (c) Magnetic field dependence of longitudinal conductivity converted from the longitudinal resistivity data. For extracting the chiral anomaly-induced conduction, a parabolic fitting is conducted to the field region with well-developed quantum oscillations so as to avoid the contributions from the size effect. The solid circles indicate the oscillation nodes.}
\label{fig3}
\end{center}\newpage
\end{figure}

The second possible scenario is the conductivity fluctuation effect. It has been discussed that spatial inhomogeneity in samples can cause distorted current path even in a Hall-bar geometry, leading to the negative MR similar to the current jetting effect \cite{CAexp5}. To demonstrate that this scenario does not apply to our case, we next present the Zn doping effect on the appearance of the negative MR.
  
When the Zn concentration $x$ is increased, the MR of {\CZA} exhibits mainly two features (Figs. 4(a) and 4(b)). One is that the negative MR is suppressed gradually as $x$ is increased in fields before the system reaches $N = 1$, where $N$ is the Landau index. Another feature is that a linear increase of MR appears around the field where $N = 1$ is realized. The positive component in high fields seems to be related to the spin-splitting of the Landau levels as discussed later. The observed suppression of the negative MR upon Zn doping is an important indication that the negative MR reflects the change of the band structure but not disorders in the samples. It is naturally expected that the Zn doping causes increasing disorders and inhomogeneity by partially replacing the Cd atoms in the lattice. The negative MR should be enhanced if the conductivity fluctuation due to disorders is the main origin.

Lastly, the gap-closing mechanism of LLLs in band-inverted materials such as topological crystalline insulator (Pb$_{1-x}$Sn$_x$)Se has been recently reported to cause negative MR in the quantum limit \cite{NMR_LLL}. This has been attributed to the enhancement in the Fermi velocity around the fields where the inverted LLLs of conduction and valence bands close the energy gap. While {\CZA} also has the band inverted nature as DSM, we note that the observed behavior of MR in the {\CZA} thin films is qualitatively different from the (Pb$_{1-x}$Sn$_x$)Se case. According to the gap-closing scenario, the MR should be positive for lower fields and then become negative only in a certain range in the quantum limit. On the other hand, in the {\CZA} thin films, the negative MR is observed rather before reaching the quantum limit, and around the quantum limit the MR turns from negative to positive. This clear contrast suggests that different origins are underlying in the present case.  

Having examined and excluded the possibilities of other scenarios, we now turn back to chiral anomaly as a possible origin for the negative MR in the {\CZA} films.  The observation of the negative MR and particularly the systematic suppression of the negative MR by Zn doping seems consistent with the expected picture of topological phase transition from the topological Dirac semimetal to a trivial insulator. On the other hand, it should be pointed out that the electron density in the present {\CZA} films is rather high and the Fermi level is likely above the Lifshitz point of the DSM phase. Because the two Dirac cones merge into a single Dirac dispersion, the chiral LLLs will not be well defined at such Fermi level and the chiral anomaly in the sense of the quantum limit picture (Fig. 1) will not be expected. On the other hand, as discussed in Ref.\cite{CA3}, the same charge pumping mechanism can arise from the non-zero Berry curvature of the band structure in the semiclassical (weak field) limit. Standing on this viewpoint, the appearance of negative MR can be expected even when the Fermi Level is above the Lifshitz point. In fact, the calculations based on the models of Weyl semimetal have shown that the Berry curvature quickly decays but remains finite even above the Lifshitz point \cite{BC1, BC2}. Therefore, the negative MR observed in the present samples presumably reflects the remnant of the non-zero Berry curvature above the Lifshitz point.

For a more quantitative evaluation of the Zn doping effect on the negative MR, we have adopted magneto-conductivity formula of the chiral anomaly derived in the semiclassical limit \cite{CA3}. The chiral-anomaly-induced conductivity was originally proposed to be proportional to the field $B$, reflecting the field dependence of the density of states of LLLs ($\varpropto eB/\hbar$) \cite{CA1}. In the diffusive limit, the field dependences of the scattering time $\tau$ and Fermi velocity $v_{\rm F}$ need to be taken into account \cite{CA2, PMR1, PMR2}, which modify the dependence as quadratic to $B$ when the inter-valley scattering is assumed as the dominant scattering process \cite{CA2, PMR1}. The same field dependence can be also derived by incorporating the Berry curvature effect in the semiclassical transport equation, where the longitudinal magneto-conductivity can be written as $\sigma_{xx} = \frac{e^{2}v_{\rm F}^{3}\tau_{\rm v}}{4 \pi^{2} \hbar E_{\rm F}^{2}} B^{2}$ ($E_{\rm F}$ is the Fermi energy) \cite{CA3}. The coefficient of the quadratic term allows an estimate of the inter-valley scattering time $\tau_{\rm v}$ or the relaxation time of the pumped charge. 

\begin{figure}
\begin{center}
\includegraphics[width=12.5cm]{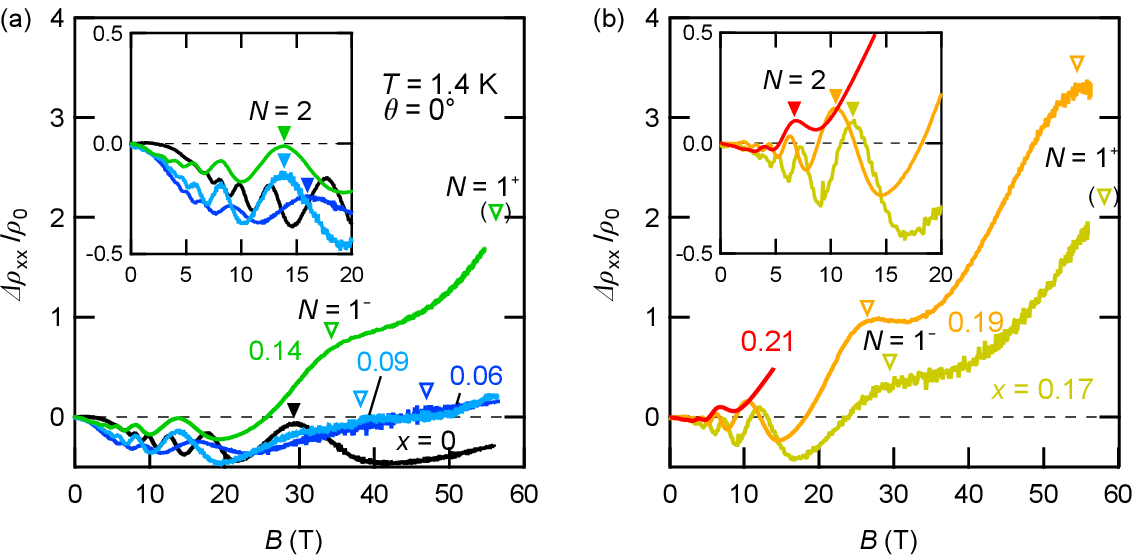}
\caption{
(color online). (a),(b) Zn concentration dependence of the longitudinal magnetoresistance (MR). Inset is the magnification of the MR curves at lower fields. The solid triangle denotes the Landau levels of $N =2$. The negative MR up to $N =2$ is gradually suppressed by Zn doping, and completely disappears at $x = 0.21$, indicating the process of the topological phase transition from DSM to a trivial insulator. In the high field region where the system reaches $N = 1$, the MR turns from negative to positive due to the spin splitting ($N = 1^{-}$ and $N = 1^{+}$ denoted by open triangles) as discussed in the main text. 
}
\label{fig4}
\end{center}
\end{figure}

In Fig. 5, we present the variation of the transport properties depending on the Zn concentration $x$. $\tau_{\rm v}$ is estimated by a parabolic fitting to the magnetoconductivity as shown in Fig. 3(c) and the Supplemental Material \cite{supplemental}, while the quantum scattering time $\tau_{\rm q}$, the effective mass $m^*$, and the Fermi velocity $v_{\rm F}$ are extracted from the quantum oscillations using the conventional Dingle expression (see \cite{supplemental} for details). 

The most striking feature here is the difference in Zn concentration dependence of the two scattering times. Reflecting the suppression of the negative MR by Zn doping in Figs. 4(a) and 4(b), $\tau_{\rm v}$ gradually decreases and finally becomes not-well defined at $x = 0.21$ where the negative MR completely disappears. In contrast, $\tau_{\rm q}$, which measures the rate of scattering events of all scattering angles, remains almost constant up to $x = 0.21$. This difference can be considered as the direct consequence of the topological phase transition from DSM to a trivial insulator. 
Reflecting the Berry curvature, the negative MR is quite sensitive to the change of the band topology. On the other hand, $\tau_{\rm q}$ probes the density of states at the Fermi level. As indicated by the small changes in $m^*$ and $v_{\rm F}$ within the present range of Fermi level, the density of states only exhibits indirect modulation through the topological phase transition. 

Early magneto-optical experiments and theoretical studies \cite{CZA2} have reported that the energy gap of {\CZA} varies linearly with $x$ (Fig. 1), and accordingly, the critical composition $x_{\rm c}$ where the band inversion disappears is estimated to be around 0.17. In recent transport measurements on bulk single crystals \cite{CZA3}, it was estimated to take place around $x = 0.38$, but analyses of only the quantum oscillations do not provide a direct probe of the band structure below the Fermi level. In contrast, our observations of the $x$-dependent negative MR clearly indicate the modulation of the band structure from DSM to trivial insulator through $x_c$. 

\color{black}
\begin{figure}
\begin{center}
\includegraphics[width=8.6cm]{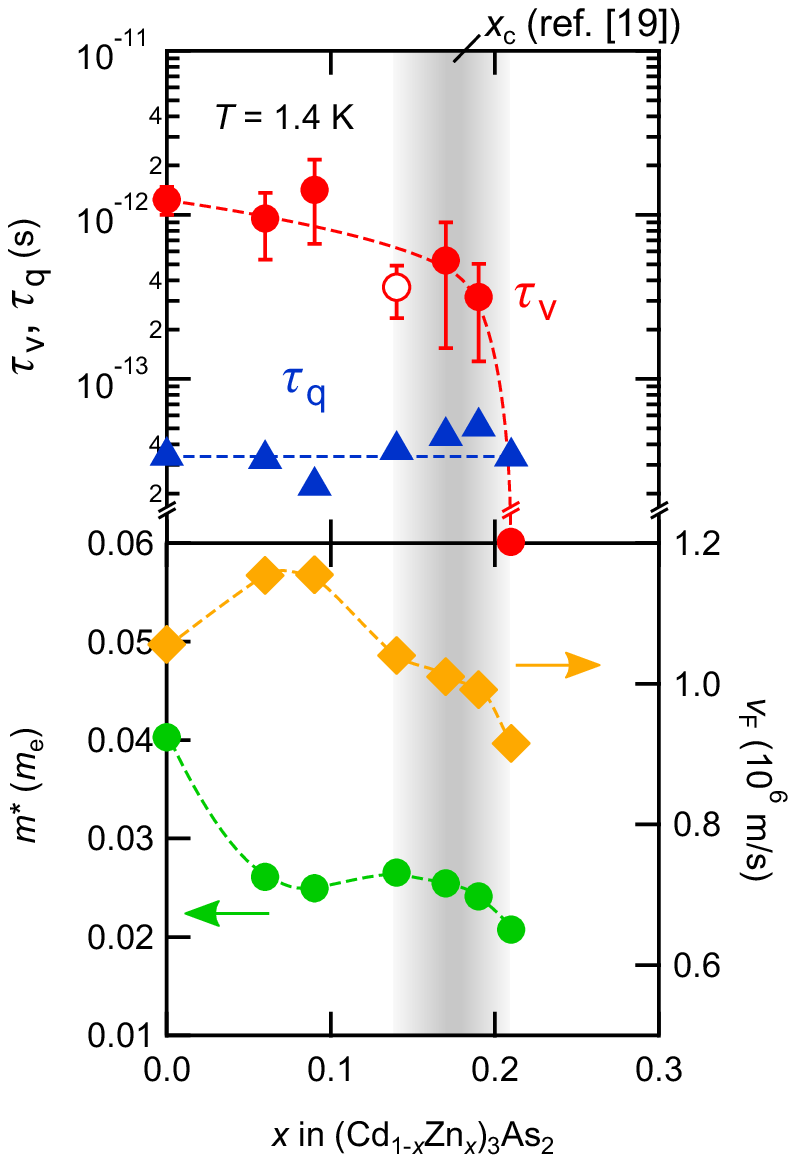}
\caption{
(color online).
Zn concentration dependence of quantum transport properties. The upper panel summarizes the intervalley scattering time $\tau_{\rm v}$ and the quantum scattering time $\tau_{\rm q}$, and the lower panel the effective mass $m^*$ and the Fermi velocity $v_{\rm F}$. The $\tau_{\rm v}$ is obtained by fitting the negative longitudinal magnetoresistance to a parabolic function, while the other parameters are estimated from the analysis of the quantum oscillations \cite{supplemental}. The systematic suppression of $\tau_{\rm v}$ contrasts with the small variation in the other parameters and it clearly reflects the modulation of the band structure from DSM towards a trivial insulator by Zn doping. The value of $\tau_{\rm v}$ for $x = 0.14$ (open circle) is underestimated due to the misalignment of the magnetic field \cite{supplemental}. 
}
\label{fig4}
\end{center}
\end{figure}

In Fig. 6, temperature dependence of MR for representative samples ($x = 0$, $0.06$, and $0.17$) is presented. The negative MR survives up to room temperature for $x = 0$, while it is clearly suppressed when approaching $x_{\rm c}$. This different robustness against thermal perturbation such as phonon scattering again indicates modulation of the band structure by Zn doping, where smaller momentum transfers are required for the inter-valley scattering due to the decreased distance between the Dirac points. 

We finally discuss the recovery of the positive MR around $N = 1$ for the $x \leqq 0.19$ samples. Similar observations have been also reported around the quantum limit in bulk {\CA} \cite{CAexp2, CAexp3, CAexp4}. It has been recently predicted that even in the regime of chiral anomaly, the spin-splitting of the doubly degenerate LLLs in DSM can yield positive linear MR \cite{PMR1, PMR2}. In our {\CZA} films, the appearance of the positive MR around $N = 1$ is indeed accompanied with the spin-splitting of the Landau levels (Figs. 4(a) and 4(b)). In Fig. 6(c), the positive MR exhibits a clear decrease in its magnitude at higher temperatures. Such a temperature dependence can be explained by blurred spin-splitting due to thermal broadening of the Landau levels, thus supporting that the spin-splitting is the underlying mechanism of the recovery of the positive MR.

\begin{figure}
\begin{center}
\includegraphics[width=12.5cm]{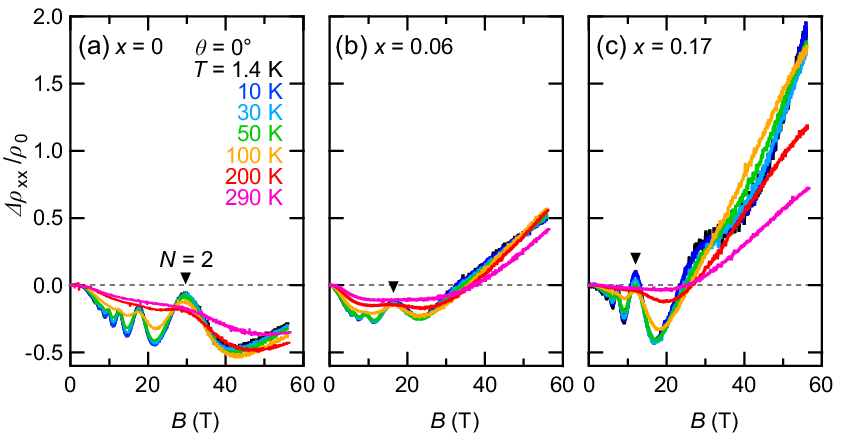}
\caption{
(color online).
Temperature dependence of longitudinal magnetoresistance (MR) for (a) $x = 0$, (b) $x = 0.06$, and (c) $x = 0.17$. The solid triangle denotes the spin degenerate Landau levels of $N = 2$. The negative MR persists up to the room temperature for $x = 0$, while it quickly suppresses for $x = 0.17$. The positive MR around $N = 1$ for $x = 0.17$ exhibits a decrease in the magnitude as the spin-splitting of the Landau levels is smeared out at higher temperatures. 
}
\label{fig4}
\end{center}
\end{figure}

In summary, we have studied magnetotransport of {\CZA} thin films up to high fields. The negative MR is systematically suppressed and finally disappears though Zn doping, indicating that the observed MR is ascribed to the chiral charge pumping and the band structure is modulated through the topological phase transition from DSM to a trivial insulator. The recovery of the positive MR at higher fields is related with the spin-splitting of the Landau levels. Our work on the systematic tuning of SOC by Zn doping in the DSM {\CA} not only reveals the manifestation of the band structure change in transport measurements, but also paves the way for realizaing more types of transport anomaly theoretically predicted in DSM \cite{Topotr3, Topotr4}. 

We thank H. Ishizuka, K. Fukushima, N. Nagaosa, M. Ueda and Y. Tokura for fruitiful discussions. This work was supported by JST CREST Grant No. JPMJCR16F1, Japan, by Grant-in-Aids for Scientific Research (C) No. JP15K05140 and Scientific Research on Innovative Areas “Topological Materials Science” No. JP16H00980 from MEXT, Japan, and by Iketani Science and Technology Foundation, Japan.

\end{document}